\documentclass[amssymb,prl,twocolumn,superscriptaddress,showpacs]{revtex4}
\usepackage{amsmath}
\usepackage{graphicx}
\usepackage{verbatim}
\usepackage{graphics}

\newcommand{\bea}{\begin{eqnarray}}
\newcommand{\eea}{\end{eqnarray}}
\newcommand{\be}{\begin{equation}}
\newcommand{\ee}{\end{equation}}

\begin{document}

\title{Mesoscopic Coulomb drag, broken detailed balance
and fluctuation relations}
\author{Rafael S\'anchez}
\affiliation{D\'epartement de Physique Th\'eorique,
Universit\'e de Gen\`eve, CH-1211 Gen\`eve 4, Switzerland}
\author{Rosa L\'opez}
\affiliation{Departament de F\'isica,
Universitat de les Illes Balears, E-07122 Palma de Mallorca, Spain}
\author{David S\'anchez}
\affiliation{Departament de F\'isica,
Universitat de les Illes Balears, E-07122 Palma de Mallorca, Spain}
\author{Markus B\"uttiker}
\affiliation{D\'epartement de Physique Th\'eorique,
Universit\'e de Gen\`eve, CH-1211 Gen\`eve 4, Switzerland}
\date{\today}

\begin{abstract}
When a biased conductor is put in proximity with an unbiased conductor a
drag current can be induced in the absence of detailed balance. This is known as the Coulomb drag effect. However, even in this 
situation far away from equilibrium where detailed balance is explicitly
broken, theory predicts that fluctuation relations are
satisfied. This surprising effect has, to date, not been confirmed
experimentally. Here we propose a system consisting of a capacitively
coupled double quantum dot where the nonlinear fluctuation relations
are verified in the absence of detailed balance.
\end{abstract}
\maketitle

\emph{Introduction}---Mesoscopic physics offers a unique laboratory to
investigate the extension of equilibrium-fluctuation dissipation theorems into the non-linear
non-equilibrium regime~\cite{heidi}. The equilibrium fluctuation-dissipation theorem and its closely related Onsager symmetry relations~\cite{casimir} are a corner stone of linear transport.
It has therefore been natural to ask whether such relations exist also
if the system is driven out of the linear transport regime. For steady state transport, fluctuation relations have been developed which relate higher order response functions to fluctuation properties of the system~\cite{heidi,david1,saito,andrieux,tobiska}. For example the current response to second order in the voltage (the second order conductance) is related 
to the voltage derivative of the noise of the system and, in
the presence of a magnetic field, to the third cumulant of the current
fluctuations at equilibrium \cite{heidi,saito}.

Clearly tests of non-equilibrium fluctuation relations are of fundamental interest. From a theoretical point of view, the task is to propose tests in which crucial relations valid at equilibrium fail in the non-linear regime and to demonstrate that, despite such a failure, fluctuation relations hold.
For instance we have suggested experiments which test fluctuation relations
for systems in the presence of a magnetic field and in a regime where the
Onsager relations are already known to fail~\cite{heidi,david2}. Such 
experimental tests are currently under way~\cite{kobayasi}.
Here we propose to test
fluctuation relations in a system where away from equilibrium we have no
detailed balance.  We consider two quantum dots in close proximity to each
other such that they interact via long range Coulomb forces. The absence of
detailed balance is manifest in a Coulomb drag~\cite{mortensen}: the
charge noise of one of the systems (the driver) drives a current through the
other unbiased system~\cite{tanatar}. Therefore,
the drag current is a direct indication that this fundamental symmetry is absent.
Nevertheless, we below demonstrate that there exist fluctuation relations. 
\begin{figure}[t]
\begin{center}
\includegraphics[width=\linewidth,clip]{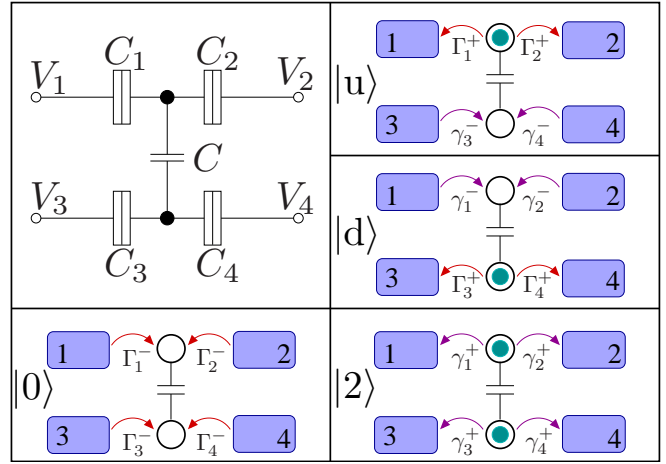}
\end{center}
\caption{\label{sys} (Color online) Sketch of two capacitively coupled quantum dots each one
attached to two different terminals.
For very large intradot charging energy, only four charge states are allowed, as depicted. 
Their dynamics is governed by the
tunneling rates $\Gamma_l^\pm$ and $\gamma_l^\pm$.}
\end{figure}

The interaction of two systems in close proximity to each other plays a role
in many important set-ups in physics. We recall here only the interaction of a
detector with a system to be measured~\cite{detection}
which also provides a test of fluctuation relations \cite{fujisawa}.
The shot noise current-current
correlations in nearby quantum dots has been measured by McClure {\it et
al.}~\cite{mcclure} and discussed theoretically~\cite{marlies,haupt}. 
Recently, reciprocity relations of two coupled conductors were proposed by
Astumian~\cite{astumian}. 
Here we emphasize that one conductor even
if unbiased can act as a gate to the other conductor.
As a consequence, the currents are not a function only of voltage differences
applied to each conductor but also depend on
potential differences of one conductor to the other one.
In an instructive work, Levchenko and Kamenev discuss the mesoscopic Coulomb
drag for two quantum point contacts in close proximity~\cite{levchenko}. In
this geometry, charging of the point contacts can be neglected and the
coupling of the two conductors is \emph{extrinsic} via the capacitance of the leads.
 
\emph{General theory}.---The probability $P(N,t)$ that $N=(N_1,\dots,N_M)$ particles are transmitted through $M$ leads during time $t$ charaterizes the statistical properties of our system. It is useful to consider the generating function which is the logarithm of the "Fourier transform" ${\cal F}(\chi)= \ln \sum_{N} P(N,t) e^{i\chi N}$ of the distribution function. Here $\chi$ is the vector of the counting fields. From the generating function all cumulants can be obtained by simple differentiation. 
The fluctuation relations are a consequence of symmetries of the generating function \cite{tobiska,andrieux}. In particular
(in the absence of a magentic field) it holds ${\cal F}(i\chi)={\cal F}(-i\chi+qV/kT)$,
which is equivalent to $P(N)=e^{NqV/kT}P(-N)$. 
Here $qV/kT$ is the
affinity vector with components given by the applied voltages $V_i$.
By expanding the current through lead $i$, ${I}_i=\langle \hat{I}_i\rangle$,
where $\hat{I}_i$ is the current operator, and the current correlations
$S_{ij}=\langle \Delta I_{i} \Delta I_{j}\rangle$,
where $\Delta I_{i}=\hat{I}_{i}-I_i$,
\begin{align}
I_i&=\sum_{j}
G_{i,j}V_j+\frac{1}{2}\sum_{jk}G_{i,jk}V_jV_k+\dots,\\
S_{ij}&=S_{ij}^{\rm eq}+\sum_k S_{ij,k}V_k+\dots, 
\end{align}
we can relate linear response current and equilibrium fluctuations
by means of the fluctuation-dissipation theorem, $S_{ij}^{\rm eq}=2kTG_{i,j}$.
The generalization to the weakly nonlinear regime reads~\cite{heidi,andrieux,saito,david1,note}
\bea\label{heidiformula}
S_{\alpha\beta,\gamma}+S_{\alpha\gamma,\beta}+S_{\beta\gamma,\alpha}=
kT(G_{\alpha,\beta\gamma}+G_{\beta,\alpha\gamma}+G_{\gamma,\alpha\beta}).
\eea
Notably, we find that these nonlinear fluctuation relations are
valid even in the absence of detailed balance.

To determine the general current-voltage characteristics and the
nonlinear fluctuations relations for two interacting conductors,
we employ the classical treatment of the Coulomb interaction that respects charge
conservation (gauge invariance).
We take the interaction to be \emph{intrinsic}, determined by the charges on the mesoscopic
conductors, and assume the leads to be metals with perfect screening. Then, 
the dynamics of the system is determined by the
sequential tunneling between states with a well defined charge occupation which obey
the master equation $\dot\rho(t)=\mathcal{M}\rho(t)$ for the occupation probabilities~\cite{cb}. 
Analogously, one can write the equation of motion 
for the generating function,
$\zeta(t)=\sum_{N} P(N,t) e^{i N\chi}$, given by 
$\dot\zeta(t)={\cal M}(\chi)\zeta(t)$. 
The cumulant generating function $\cal F$ is given by
the eigenvalue of $\mathcal{M}(\chi)$
that develops adiabatically from zero with small $\chi$~\cite{bagrets}.
Generally, the explicit expression for $\mathcal{F}$
is difficult to handle, so in practice it is more convenient to calculate the cumulants
recursively order by order~\cite{fcs,flindt}. Thus,
from coefficients $c_{\{l_k\}}$ of 
the expansion 
$\mathcal{F}=\sum_{\{l_k\}} c_{\{l_k\}} (e^{i\chi_{1}}-1)^{l_1}\dots(e^{i\chi_{M}}-1)^{l_M}$,
we obtain the current-current correlations up to any order~\cite{fcs}; e.g., the
current, $I_{i}= q\sum_{\{l_k\}}c_{\{l_k\}}\delta_{l_i,1}\delta_{\sum l_k,1}$,
the zero frequency noise,
$S_{ii}=qI_i+2q^2\sum_{\{l_k\}}c_{\{l_k\}}\delta_{l_i,2}\delta_{\sum l_k,2}$,
and the cross-correlations
$S_{ij}=q^2 \sum_{\{l_k\}}c_{\{l_k\}}\delta_{l_i,1}\delta_{l_j,1}\delta_{\sum l_k,2}$,
where ${\{l_k\}}={\{l_1\dots l_M}\}$ and $q$ is the electron charge. 

\emph{Drag current and fluctuation relations}.---In the following we explicitly show,
using the previous formalism,
the fulfillment of the fluctuation relations in a nonequilibrium system where
detailed balance is broken. We consider
two capacitively coupled two-terminal quantum dots (see Fig.~\ref{sys}) with large intradot charging energy.
The interdot coupling is described with a capacitance $C$. Hence,
the dynamics is characterized by four charge states: the empty state
$|0\rangle=|00\rangle$, the singly occupied states $|{\rm u}\rangle=|10\rangle$ and
$|{\rm d}\rangle=|01\rangle$, and the doubly occupied state $|2\rangle=|11\rangle$. Quite generally, the tunneling amplitudes are energy dependent.
Therefore, we distinguish $\Gamma_l$, which denotes a tunneling process through barrier
$l=1,\ldots,4$ when the system is empty,
and $\gamma_l$, which corresponds to tunneling when the coupled dot is already occupied.
This is the {\it minimal} charge model that 
manifests violation of detailed balance leading to drag currents.
Detailed balance is broken when the probability to transfer one charge from left to
right differs from the reverse process (from right to left).
For instance, an electron is transported from left to right in the drag system by the
sequence
$|0\rangle{\rightarrow}|{\rm u}\rangle{\rightarrow}|2\rangle{\rightarrow}|{\rm d}\rangle{\rightarrow}|0\rangle$ 
with a probability $\propto\Gamma_{\!1}\gamma_2$ whereas 
the probability to transport it from right to left is
$\propto \gamma_1\Gamma_{\!2}$. Clearly, both probabilities
differ and a nontrivial current, the drag current $I_{\rm drag}\propto \Gamma_{\!1}\gamma_2-\gamma_1\Gamma_{\!2}$,
will be generated. We need that
{\it (i)} both empty and doubly occupied states are taken into account
and {\it (ii)} the tunneling rates depend on the charge state.
Thus, a model with three charge states only
($|{\rm u}\rangle$, $|{\rm d}\rangle$ and $|0\rangle$ or $|2\rangle$) cannot break the
detailed balance and the drag effect is
absent. The biased dot then acts merely as a fluctuating gate on the other dot.

For the system depicted in Fig~\ref{sys}, writing $\zeta=(\zeta_{0},\zeta_{u},\zeta_{d},\zeta_2)$,
the equation $\dot\zeta={\cal M}(\chi)\zeta$ becomes,
\begin{eqnarray}\label{MLR}
\displaystyle
{
{\cal M}\!=\!\!
\left(\begin{array}{cccc}
-\Gamma_{\!u}^--\Gamma_{\!d}^- &\tilde\Gamma_{\!u}^+ & \tilde\Gamma_{\!d}^+ & 0 \\
\tilde\Gamma_{\!u}^- &  -\Gamma_{\!u}^+-\gamma_{d}^- & 0 & \tilde\gamma_{d}^+ \\
\tilde\Gamma_{\!d}^- & 0 & -\gamma_{u}^--\Gamma_{\!d}^+ & \tilde\gamma_{u}^+ \\
0 & \tilde\gamma_{d}^- & \tilde\gamma_{u}^- & -\gamma_{u}^+-\gamma_{d}^+ \\
\end{array}  \right),
}
\end{eqnarray}
where $\tilde\Gamma_{\alpha}^\pm=\sum_{l\in\alpha}e^{\pm
i\chi_l}\Gamma_{l}^\pm$, and
$\tilde\gamma_{\alpha}^\pm=\sum_{l\in\alpha}e^{\pm i\chi_l}\gamma_{l}^\pm$,
$u=\{1,2\}$ and $d=\{3,4\}$.
The tunneling rates \emph{in}(-) and
\emph{out}(+) of the dot read
$\Gamma_{l}^\pm=\Gamma_{\!l} f^\pm_{l0}$ 
and $\gamma_{l}^\pm=\gamma_l f^\pm_{l1}$ with $f^\pm_{ln}=f^\pm(\mu_{ln}-qV_l)$ ($n=0,1$).
Here, $f^+(\varepsilon)=1-f(\varepsilon)$ and
$f^-(\varepsilon)=f(\varepsilon)$ denote the hole and electron Fermi functions,
respectively. The effective level of dot $\alpha$ with bare level $\varepsilon_\alpha$
when dot $\beta\neq\alpha$ is uncharged ($n=0$) is
$\mu_{\alpha0}=\varepsilon_\alpha+[q^2C_{\Sigma\alpha}/2
+q(C_{\Sigma\beta}\sum_{l\in\alpha}C_lV_l+C\sum_{l\in\beta}C_lV_l)]/C\tilde C$,
where $C_l$ is the capacitance of the $l$th barrier,
$C_{\Sigma\alpha}=\sum_{l\in\alpha}C_l+C$
and $\tilde C=(C_{\Sigma\rm u}C_{\Sigma\rm d}-C^2)/C$.
In the charged case ($n=1$), we find $\mu_{\alpha1}=\mu_{\alpha0}+E_{\rm C}$ with 
$E_{\rm C}=2q^2/\tilde C$ the energy needed to add a second electron. 
\begin{figure}[t]
\begin{center}
\includegraphics[width=0.493\linewidth,clip]{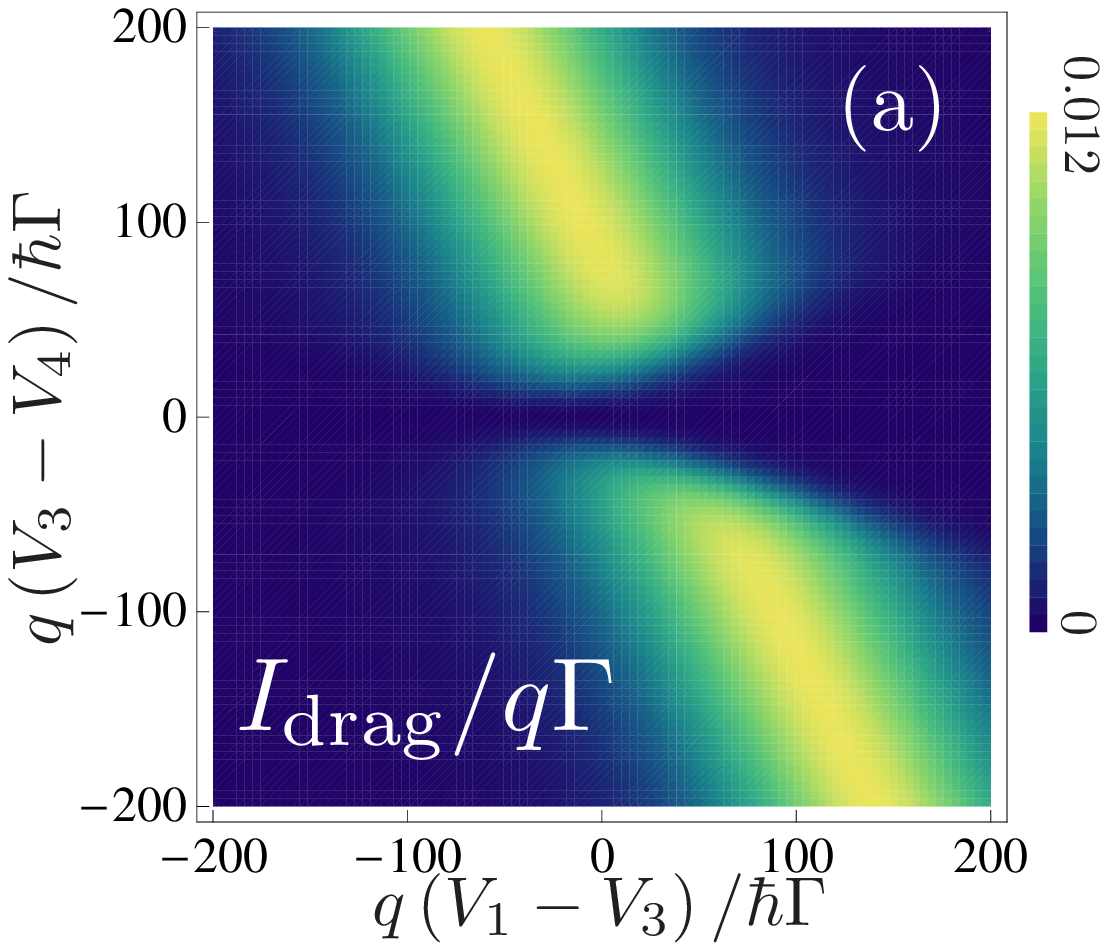}
\includegraphics[width=0.493\linewidth,clip]{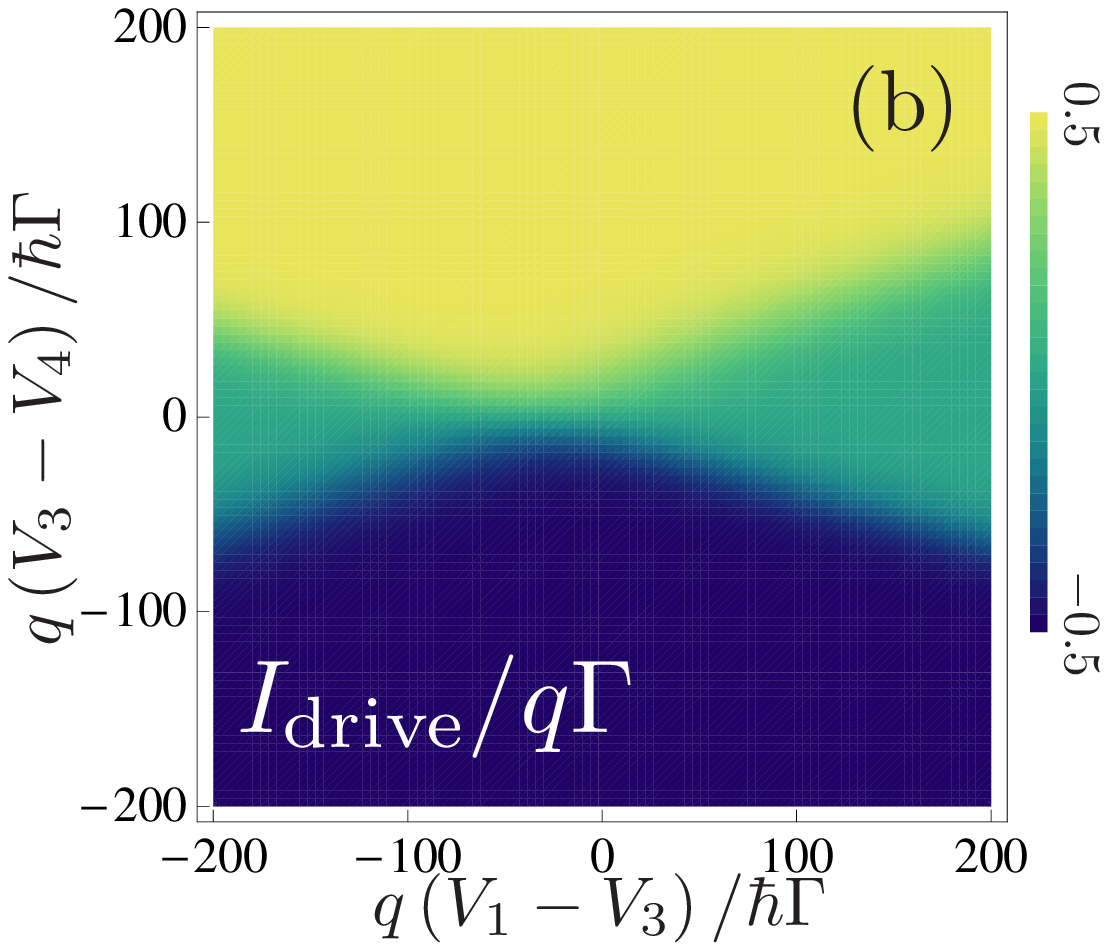}
\includegraphics[width=0.493\linewidth,clip]{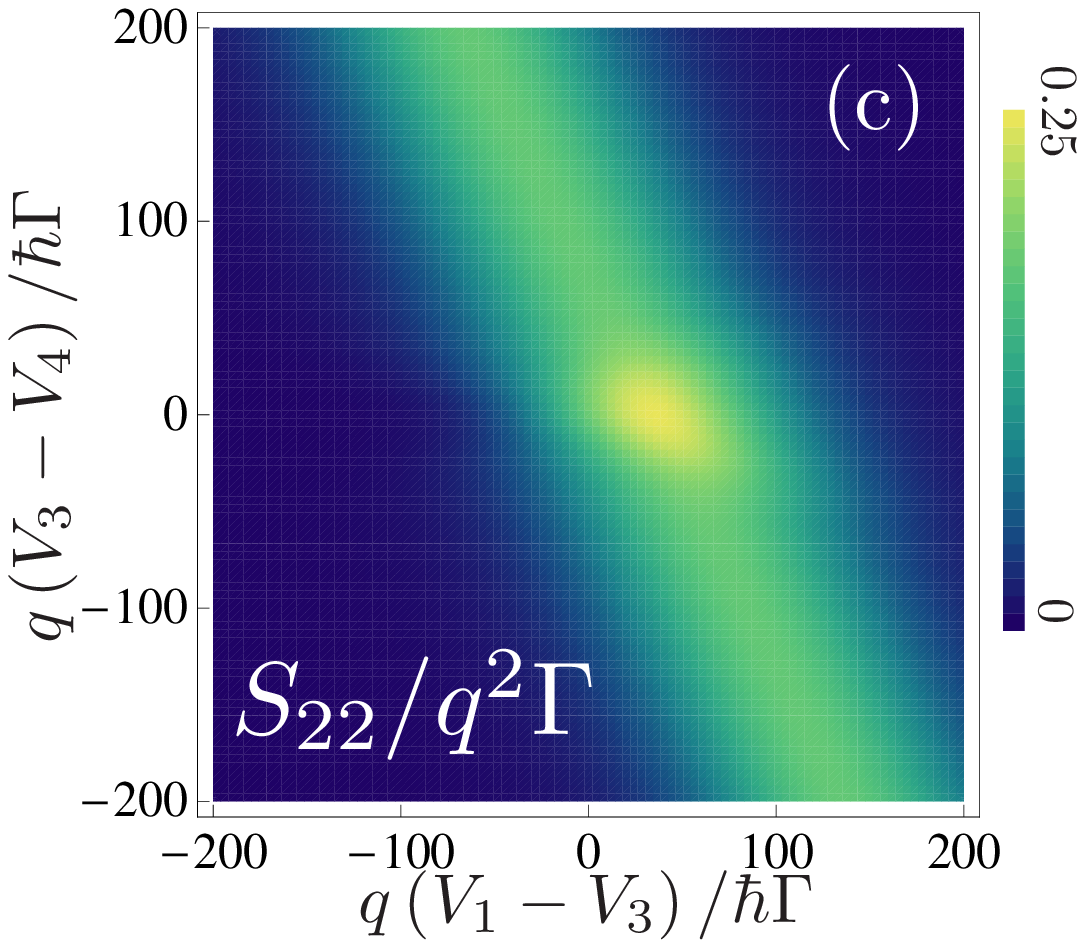}
\includegraphics[width=0.493\linewidth,clip]{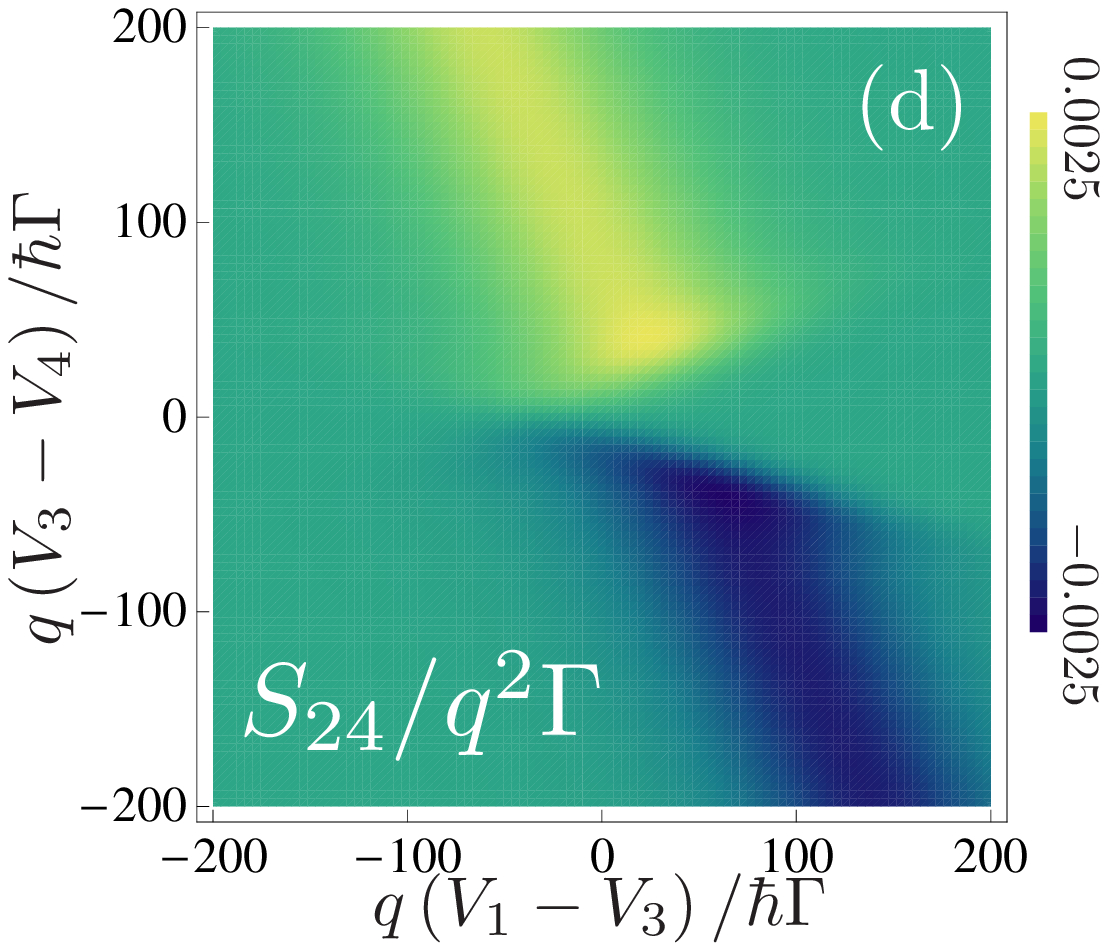}
\end{center}
\caption{\label{isv} (Color online) Voltage dependence of (a) the drag current  $I_{\rm drag}$ through
the upper dot ($V_1=V_2$) 
(b) the drive current $I_{\rm drive}$ for the lower dot, (c)
the current-current correlations in the drive system, $S_{22}=\langle
\Delta I_{2}\Delta I_{2}\rangle$ and (d) the cross-correlation between currents at different conductors,
$S_{24}=\langle \Delta I_2\Delta I_4
\rangle$, for the drag configuration, $V_1=V_2$. 
Parameters: $\Gamma_{\!i}=\gamma_i=\Gamma$, except $\gamma_{1}=0.1\Gamma$, $kT=5\hbar\Gamma$, $q^2/C_i=20\hbar\Gamma$, $q^2/C=50\hbar\Gamma$, $\varepsilon_{\rm u}=\varepsilon_{\rm d}=0$.
}
\end{figure}

We now investigate the drag current, for which 
we take the upper subsystem
as the drag circuit ($V_1=V_2$) and the lower one as the driver. 
Then, $I_1=-I_2=I_{\rm drag}$ and we find
\be
\label{dragcurr}
I_{\rm drag}=\frac{q\left(\gamma_1\Gamma_{\!2}-\gamma_2\Gamma_{\!1}\right) \Gamma_{\!d}\gamma_d
\sum_{\sigma=\{+,-\}} \sigma f_{10}^\sigma f_{11}^{-\sigma} g_0^{-\sigma} g^\sigma_1}{\Gamma_{\!u}\Gamma_{\!d}\left(\gamma_u h^+ +\gamma_d
k^+\right)+\gamma_u\gamma_d\left(\Gamma_{\!u} h^-+\Gamma_{\!d}
k^-\right)}\,,
\ee 
where $\Gamma_{\alpha}=\sum_{l\in\alpha}\Gamma_{\!l}$,
$\gamma_{\alpha}=\sum_{l\in\alpha}\gamma_{l}$,
$h^{\pm}=f^\pm_{11}\pm g^-_{0(1)} \left(f^-_{11}-f^-_{10}\right)$, and
$k^{\pm}=g^\pm_{1}\pm  f^-_{10(11)} \left(g^-_1-g^-_0\right)$.
$g_0^{\pm}=(\Gamma_{\!3} f_{30}^{\pm}+\Gamma_{\!4} f_{40}^{\pm})/\Gamma_{\! d}$
and $g_1^{\pm}=(\gamma_3 f_{31}^{\pm}+\gamma_4 f_{41}^{\pm})/\gamma_d$
are nonequilibrium distribution functions.

When the drive voltage $V_3-V_4$ is small,
detailed balance must be broken also in the drive circuit in order
to have a linear $I_{\rm drag}$:
$G_{2,4}\propto (\gamma_1\Gamma_{\!2}-\gamma_2\Gamma_{\!1})(\gamma_4\Gamma_{\!3}-\gamma_3\Gamma_{\!4})$.
Therefore, asymmetry in both the drag and the drive systems is required for a nonzero linear drag current. 
Moreover, we get
$G_{2,4}^{(1)}=G_{4,2}^{(1)}$,
satisfying the Onsager-Casimir reciprocity relations~\cite{casimir}. Note that if
the drive conductor is also unbiased ($V_{3}=V_4$), equilibrium fluctuations
are expectedly not enough to induce a net current.
This can be seen in Fig. \ref{isv}(a). For low voltages
there is a Coulomb gap where transport is not allowed. This result also demonstrates
that the voltage difference between the two subsystems $V_1-V_3$ plays a
crucial role, affecting the dynamics: 
In this case, one of the conductors acts as a gate on the other one.
The gate effect of the drag circuit onto the driver is shown
in Fig. \ref{isv}(b), where we obtain a typical Coulomb blockade stability diagram
for the drive current $I_3=-I_4=I_{\rm drive}$. 

It is worth noticing that, at high enough drive bias, $I_{\rm drag}$ is suppressed since
the interdot capacitance brings the dot states outside the transport
window. Then, the drag current peaks at an optimal
value of $V_{1}-V_3$ and vanishes away from it.  
On the other hand, at very low temperature $I_{\rm drag}$ is finite only within a
voltage range defined by $\mu_{10}<qV_1<\mu_{11}$ and ${\rm min}\{qV_3,qV_4\}<\mu_{10},\mu_{11}<{\rm max}\{qV_3,qV_4\}$. As expected, the drag current increases with $C$, but the voltage window where $I_{\rm drag}$ is observable becomes narrower. Then, for large coupling the drag circuit effectively induces dynamical channel blockade~\cite{belzig}
in the driver and, eventually, the drive current shows electron bunching.

If the drive system is symmetric, the sign of $I_{\rm drag}$
depends on the asymmetry factor ($\Gamma_{\!1}\gamma_2-\gamma_1\Gamma_{\!2}$)
due to the competition of 
processes transferring an electron in each direction,
independently of the direction of $I_{\rm drive}$.
These two contributions have been detected separately in coupled double dot systems
in the cotunneling regime giving rise to bidirectional drag~\cite{shinkai}.
Note that the asymmetry of the drag system can be enough to get a {\it negative} drag.

We now investigate the nonlinear fluctuation relations for our
system. We first analyze the occurrence of $I_{\rm drag}$ and the
current cross-correlations $S_{ij}$
for different conductors (e.g., $i=\{1,2\}$ and $j=\{3,4\}$). The observation of
drag current in one conductor requires the occurrence of correlated tunneling
events between the two dots involving the states
$|0\rangle$ and $|2\rangle$. These correlated events lead to
finite cross-correlations. 
This would not be the case for a model that includes only three
charge states. Our minimal model of four charge states does
generate correlations between the currents through the two dots. For
example, at equilibrium, the fluctuation-dissipation theorem relates
the linear {\em drag} current to the  equilibrium cross-correlations for
{\em different} conductors,
$G_{2,4}=S_{24}^{\rm eq}/2kT$. 
Similarly to $I_{\rm drag}$, if both conductors are symmetric, i.e.,
$\gamma_1\Gamma_{\!2}=\gamma_2\Gamma_{\!1}$ and $\gamma_3\Gamma_{\!4}=\gamma_4\Gamma_{\!3}$,
$S_{ij}$ vanishes to
first order in a voltage expansion. Figure \ref{isv}(d) shows 
that the cross-correlation between the drag and drive currents
is finite only when there
is a drag current flowing in the upper conductor. In
general, the sign of the cross-correlations is not determined by the
direction of the averaged currents~\cite{mcclure}. However, 
in our case, the cross-correlations are positive whenever the two currents flow
in the same direction, and negative when they are opposite. Interestingly,
$I_{\rm drag}$ can present negative excess noise, i.e., the noise $S_{22}$
{\em decreases} in the presence of drag, as shown in Fig. \ref{isv}(c).
$S_{22}$ reaches its
maximal value when the effective upper dot level is aligned
with the Fermi level~\cite{lesovik}.
\begin{figure}[t]
\begin{center}
\includegraphics[width=\linewidth,clip]{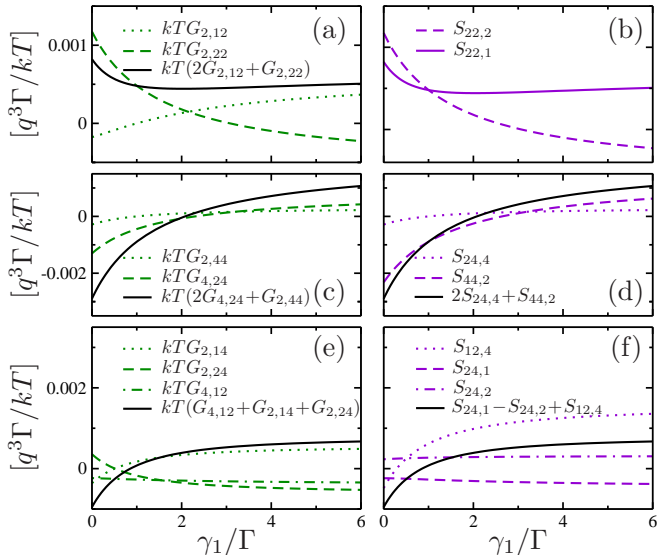}
\end{center}
\caption{\label{fig:fluct} (Color online) Verification of the
non-equlibrium fluctuation relations. Left panel: nonlinear
conductances for (a) Eq.~(\ref{Salfaalfa}), 
(c) 
and (e) Eq.~(\ref{Salfabetabeta}). Right panel: noise susceptibilities
for the corresponding relations in the left.
Same parameters as in Fig.~\ref{isv}, except $\gamma_4=0.1\Gamma$.}
\end{figure}

Finally, we explicitly check that these fluctuation relations~\cite{heidi,david1,saito,andrieux}
hold even for our system in which detailed balance is violated.
Charge conservation in each subsystem implies $I_\alpha=-I_{\bar\alpha}$ and $S_{\alpha\alpha}=S_{\bar\alpha\bar\alpha}=-S_{\alpha\bar\alpha}$
for two different terminals in the same conductor.
Then, from Eq.~(\ref{heidiformula}) we derive the nonlinear fluctuation relations involving
terminals of the same conductor, and rewrite them as
\begin{align}
\label{Salfaalfa}S_{\alpha\alpha,\alpha}&=kT G_{\alpha,\alpha\alpha}, \nonumber\\[-2.5 mm]
&\\[-2.5 mm]
S_{\alpha\alpha,\bar\alpha}&=-kTG_{\alpha,\bar\alpha\bar\alpha}=kT(2G_{\alpha,\alpha\bar\alpha}+G_{\alpha,\alpha\alpha}),\nonumber
\end{align}
with $G_{\alpha,\alpha\alpha}+G_{\alpha,\bar\alpha\bar\alpha}+2G_{\alpha,\alpha\bar\alpha}=0$. 
In Fig.~\ref{fig:fluct}(a) and (b) we explicitly check that these fluctuation relations hold
despite broken detailed balance.
The relations including derivatives of current cross-correlations
at different conductors, $\alpha$ and $\beta$, read
\begin{align}
\label{Salfabetabeta}
& 2S_{\alpha\beta,\beta}+S_{\beta\beta,\alpha}=kT(G_{\alpha,\beta\beta}+2G_{\beta,\beta\alpha}),\nonumber\\[-2.5 mm]
&\\[-2.5 mm]
\label{Salfabetabaralfa}&
S_{\alpha\beta,\bar\alpha}-S_{\alpha\beta,\alpha}+S_{\bar\alpha\alpha,\beta}
=kT(G_{\beta,\alpha\bar\alpha}+G_{\alpha,\bar\alpha\beta}-G_{\alpha,\alpha\beta}),\nonumber
\end{align}
with $\sum_{\alpha,\beta}G_{\gamma,\alpha\beta}=0$.
Eq. (\ref{Salfabetabeta}) is verified in Figs.~\ref{fig:fluct}(c-f).
It is important to
realize here that full access to the fluctuation relations is
{\em only} possible in the presence of drag current [see Fig.~\ref{isv}(a) and (c)].
In other words, only when detailed balance is broken and a drag current appears 
\emph{are all fluctuation relations nontrivially verified}. In contrast, 
the absence of drag, i.e.
$G_{\alpha,\beta\beta}=G_{\alpha,\beta\bar\beta}=0$, implies
$S_{\alpha\beta,\gamma}=0$, for any terminal $\gamma$, in which case the fluctuation relations (\ref{Salfabetabeta}) 
are simply reduced to the relation $S_{\alpha\alpha,\beta}=2kTG_{\alpha,\alpha\beta}$, with $G_{\alpha,\alpha\beta}=-G_{\alpha,\bar\alpha\beta}$.

{\it Conclusions}.---In summary,
we have proposed a geometry of two conductors put in proximity 
interacting via long-range Coulomb forces to test fluctuation
relations in the non-linear transport regime. This system exhibits a drag current as a direct consequence of the 
absence of detailed balance. Our main findings are {\it (i)} the general expression
for the current-voltage characteristic of two interacting conductors, and {\it (ii)} the
verification of the fluctuation relations in a nonequilibrium
system when detailed balance is broken. 
Our proposal motivates new experiments to
test the fluctuation relations away from equilibrium when detailed balance does not hold.

{\it Acknowledgements.}---We thank M. Moskalets, H. F\"orster, M.G. Vavilov and
C. Flindt for instructive discussions. This work was supported by MaNEP,
the Swiss NSF and the strep project SUBTLE. R.L. and D.S acknowledge
support from MICINN Grant FIS2008-00781. 

\end{document}